\documentclass[preprint2]{aastex}

\shorttitle{Pixel Analysis of NGC 5195}

\shortauthors{Lee et al.}
\def\simlt{\lower.5ex\hbox{$\; \buildrel < \over \sim \;$}}
\def\simgt{\lower.5ex\hbox{$\; \buildrel > \over \sim \;$}}

\begin{document}

\title{WHAT DETERMINES THE SIZES OF RED EARLY-TYPE GALAXIES?}

\author{Joon Hyeop Lee$^{1}$, Minjin Kim$^{1,2}$, Chang Hee Ree$^{1}$, Sang Chul Kim$^{1}$, Jong Chul Lee$^{1}$, Hye-Ran Lee$^{1,3}$, Hyunjin Jeong$^{1}$, Kwang-Il Seon$^{1}$, Jaemann Kyeong$^{1}$, Kyuseok Oh$^{4}$}
\affil{$^{1}$Korea Astronomy and Space Science Institute, Daejeon 305-348, Korea\\
$^{2}$The Observatories of the Carnegie Institution for Science, 813 Santa Barbara Street, Pasadena, CA 91101, USA\\
$^{3}$University of Science and Technology, Daejeon 305-350, Korea\\
$^{4}$Department of Astronomy, Yonsei University, Seoul 120-749, Korea}

\email{jhl@kasi.re.kr}

\begin{abstract}
The sizes of galaxies are known to be closely related with their masses, luminosities, redshifts and morphologies. However, when we fix these quantities and morphology, we still find large dispersions in the galaxy size distribution. We investigate the origin of these dispersions for red early-type galaxies, using two SDSS-based catalogs.
We find that the sizes of faint galaxies ($\log\, (M_{\textrm{\scriptsize dyn}}/M_{\odot})\simlt10.3$ or $^{0.1}M_{r}\simgt-19.5$, where $^{0.1}M_{r}$ is the $r$-band absolute magnitude, k-corrected to z = 0.1) are affected more significantly by luminosity, while the sizes of bright galaxies ($\log\, (M_{\textrm{\scriptsize dyn}}/M_{\odot})\simgt11.4$ or $^{0.1}M_{r}\simlt-21.4$) are by dynamical mass.
\emph{At fixed mass and luminosity}, the sizes of low-mass galaxies ($\log\, (M_{\textrm{\scriptsize dyn}}/M_{\odot})\sim10.45$ and  $^{0.1}M_{r}\sim-19.8$) are relatively less sensitive to their colors, color gradients and axis ratios. On the other hand, the sizes of intermediate-mass ($\log\, (M_{\textrm{\scriptsize dyn}}/M_{\odot})\sim10.85$ and  $^{0.1}M_{r}\sim-20.4$) and high-mass ($\log\, (M_{\textrm{\scriptsize dyn}}/M_{\odot})\sim11.25$ and  $^{0.1}M_{r}\sim-21.0$) galaxies significantly depend on those parameters, in the sense that larger red early-type galaxies have bluer colors, more negative color gradients (bluer outskirts) and smaller axis ratios.
These results indicate that the sizes of intermediate- and high-mass red early-type galaxies are significantly affected by their recent minor mergers or rotations, whereas the sizes of low-mass red early-type galaxies are affected by some other mechanisms. Major dry mergers also seem to have influenced on the size growth of high-mass red early-type galaxies.
\end{abstract}


\section{INTRODUCTION}

The sizes of galaxies are an important constraint on the galaxy formation and evolution scenarios.
As reasonably expected, more massive (or luminous) galaxies tend to have larger sizes, which is known to be the mass-size (luminosity-size) relation of galaxies \citep[e.g.,][]{tru04}.
From many recent studies, the mass-size or luminosity-size relation of galaxies depends on galaxy morphology and redshift, but not significantly on local environment \citep{mci05,guo09,mal10,nai10}. The cosmic evolution of those relations is an important issue in which many researchers are interested recently, because the sizes of early-type galaxies at high redshifts are too small, compared to their counterparts at low redshifts \citep{fan10,cas11,blu12}.
Several studies suggested that minor mergers may be the main driver of the galaxy size growth \citep[e.g.,][]{coo12,pat12,oog12}, while some other studies argued that major mergers may play an important role, too \citep[e.g.,][]{pap12,hue12}.

When we focus on the `sizes' of galaxies, the current knowledge about the mass-size (or luminosity-size) relation of galaxies can be re-expressed: the sizes of galaxies are correlated with their \emph{masses, luminosities, redshifts and morphologies}. However, if we fix these quantities and morphology, then do they have the same sizes?
The answer seems to be NO.
It is not sufficient to have only these 4 factors to explain the size distribution of galaxies. Even the galaxies with the same morphologies and at similar redshifts show large dispersions in their mass-size or luminosity-size relation \citep[e.g.,][]{nai10}. For example, early-type galaxies at fixed mass or fixed luminosity show size difference by up to several times.

The purpose of this letter is to investigate the origin of the dispersion in the size distribution of galaxies with fixed morphology. Here we focus on red early-type galaxies, because their sizes and velocity dispersions are more reliably estimated than those of blue late-type galaxies: sizes less affected by short-lived young stars and velocity dispersions less contaminated by rotational or transient motions.
We exclude `blue' early-type galaxies, because they have properties very different from those of red early-type galaxies \citep[e.g.,][]{lee06,lee08}.

The outline of this paper is as follows. In Section 2, we briefly describe our data sets and sample selection. The basic mass-size and luminosity-size relations are inspected in Section 3. Using mass-and-luminosity-fixed samples, we investigate the correlations between sizes and several quantities in Section 4. The discussion and conclusion are provided in Section 5.
Throughout this paper, we adopt the cosmological parameters: $h=0.7$, $\Omega_{\Lambda}=0.7$, and $\Omega_{M}=0.3$. All magnitudes are in the AB system.

\section{DATA AND RED EARLY-TYPE GALAXY SAMPLE}

For the purpose of this letter, some data sets with reliably estimated physical quantities are necessary.
We use two data sets:
\emph{Korea Institute for Advanced Study Value-Added Galaxy Catalog} \citep[KIAS-VAGC;][]{cho10} and \emph{Improved and Quality-Assessed Emission and Absorption Line Measurements in SDSS Galaxies} \citep[OSSY Catalog;][]{oh11}, both of which are based on the Sloan Digital Sky Survey Data Release 7 \citep[SDSS DR7;][]{aba09}.
The KIAS-VAGC provides many valuable morphological, photometric and structural parameters for $\sim$ 58,000 SDSS galaxies; including color gradients and axis ratios of galaxies.
The color gradient is defined as the difference in $(g-i)$ color between the region at $0.5 R_{\textrm \scriptsize pet} < R < R_{\textrm \scriptsize pet}$ \footnote{$R_{\textrm \scriptsize pet}=$ Petrosian radius.} and the region at $R < 0.5 R_{\textrm \scriptsize pet}$ (positive $\Delta(g-i)$ for blue center) and the axis ratio was estimated using ellipsoidal fitting \citep{cho07}, which were corrected for the inclination and the seeing effects. The morphologies of galaxies were classified in the 3-dimensional parameter space of color, color gradient and light concentration \citep{par05}.~\footnote{This returns at least $90\%$ completeness and reliability for the faintest galaxies ($\sim17.5$ AB mag in the $r$ band) in the sample \citep{par05,cho10}.} For 83,292 galaxies in the `trouble zones', they performed visual classifications to determine their morphologies.

The OSSY Catalog provides accurate measurements of absorption and emission lines for the entire SDSS DR7 galaxies ($z<0.2$) using a very elaborate way of line strength measurement techniques of \citet{sar06}, but we simply use the stellar velocity dispersion information in it, to estimate the dynamical masses of our sample galaxies.
For reliability, we use only the galaxies with the stellar velocity dispersions of $70\le\sigma_v<500$ km s$^{-1}$ and with the signal-to-noise (S/N) ratios $\ge20$.
These criteria reject $<5\%$ of red early-type galaxies at $\sigma_v\ge200$ km s$^{-1}$, whereas $>70\%$ at $\sigma_v\le100$ km s$^{-1}$.

Since the redshift range of the sample in this letter is relatively narrow ($0.04<z<0.10$), we can minimize the effect of cosmic evolution of galaxy sizes. Thus, in our sample, the sizes of galaxies are expected to be mainly correlated with their masses, luminosities and morphologies. To build the sample of red early-type galaxies, we selected early-type galaxies within a tight color versus dynamical mass relation. Using the linear least squares fit, the red sequence was defined to be $^{0.1}(u-r)=0.176\, \log\, (M_{\textrm{\scriptsize dyn}}/M_{\odot})+0.854$, where $^{0.1}(u-r)$ is the $u-r$ color $k$-corrected to the $z=0.1$ redshift \footnote{Similarly, $^{0.1}M_r$ is the $r$-band absolute magnitude, $k$-corrected to $z=0.1$.} and $M_{\textrm{\scriptsize dyn}}$ is the dynamical mass of a galaxy. The dynamical mass was estimated using the following equation:
\begin{equation}
M_{\textrm{\scriptsize dyn}} = \frac{2R{\sigma_v}^2}{G},
\end{equation}
where $R$ is the virial radius and $G$ is the gravitational constant. In our estimation, the Petrosian radius was used for the $R$ value and the $\sigma_v$ value was corrected for the spectroscopic fiber aperture effect, adopting the method of \citet{jor95}.
The red galaxies were selected to be within $\pm0.25$ mag from the red sequence colors, which yields 53,450 red early-type galaxies.

\section{MASS-SIZE-LUMINOSITY RELATIONS}

\begin{figure}[!t]
\plotone{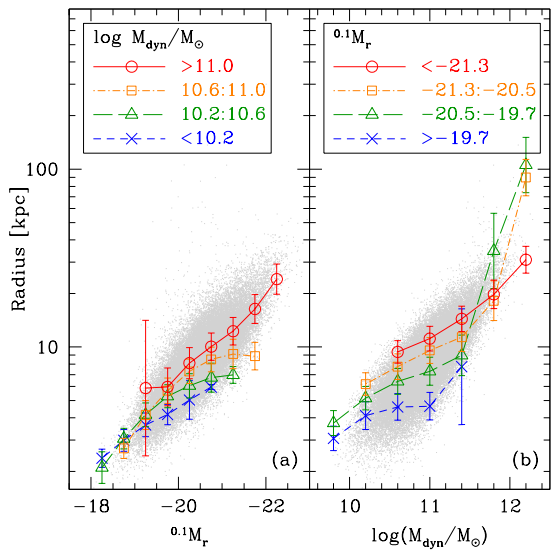}
\caption{ (a) Size versus luminosity relation of 53,450 red early-type galaxies. The median value and sample inter-quartile range (SIQR) at each luminosity bin are overlaid for different dynamical mass ranges. (b) Size versus dynamical mass relation of the same galaxies. The median value and SIQR at each log mass bin are overlaid for different luminosity ranges.
\label{msrel}}
\end{figure}

First, we compared the effects of mass and luminosity on the sizes of red early-type galaxies in Figure~\ref{msrel}.
Each of mass and luminosity is tightly correlated with size, but none of them solely dominates the sizes of the sample galaxies.
In Figure~\ref{msrel}(a), the sizes of the galaxies for a given luminosity show large deviations, which can be ascribed by difference in their dynamical masses. In a similar way, the galaxy sizes are not uniquely correlated with their dynamical masses, but show large deviations due to difference in luminosity (Figure~\ref{msrel}(b)).

However, from the comparison between Figure~\ref{msrel}(a) and Figure~\ref{msrel}(b), we can disentangle the dependency on mass from the dependency on luminosity. In Figure~\ref{msrel}(a), the dynamical mass has almost no influence on the sizes of the galaxies at the faint end ($^{0.1}M_{r}\simgt-19.5$, averagely corresponding to $\log\, (M_{\textrm{\scriptsize dyn}}/M_{\odot})\simlt10.3$), while the sizes of the galaxies at the bright end clearly depend on their dynamical masses.
On the other hand, in Figure~\ref{msrel}(b), luminosity does not affect the sizes of galaxies at the bright end ($\log(M_{\textrm{\scriptsize dyn}}/M_{\odot})\simgt 11.4$, corresponding to $^{0.1}M_{r}\simlt-21.4$), while the sizes of galaxies at the faint end are significantly affected by their luminosities.
In other words, the sizes of the faint (low-mass) galaxies are correlated with their luminosities rather than masses, whereas those of the bright (massive) galaxies are correlated with their masses rather than luminosities.

The trends in Figure~\ref{msrel} are related with the fundamental plane, the tight correlation between the effective radii, average surface brightnesses and central velocity dispersions of early-type galaxies \citep{dre87}, since luminosity is a function of radius and surface brightness, while dynamical mass is a function of radius and velocity dispersion. We checked the dependence of the sizes on the velocity dispersions, finding that $\sigma_v$ is almost independent of $R$ at the bright end, but that a weak anti-correlation between $\sigma_v$ and $R$ is found at the faint end. This means that adding $\sigma_v$ as the third parameter slightly decreases the scatter at the faint end in Figure~\ref{msrel}(a), although it is not sufficient to explain the entire scatter.

We tested the possible effect of the measurement errors for brightest cluster galaxies (BCGs) contaminated by intracluster light. This problem is not significant at low redshifts, because the SDSS spectroscopic data is incomplete for bright objects ($r<15$ mag), but some galaxies near to $z\sim0.1$ in our sample may be BCGs.
To check how different behaviors BCGs show from those of non-BCGs, we used the MaxBCG catalog \citep{koe07}. Among our 53,450 red early-type galaxies, 26 galaxies are identified to be BCGs, all of which exist at $z>0.9$. We confirmed that those BCGs agree with the relations in Figure~\ref{msrel}, which shows that the effect of measurement errors for BCGs is not significant in our study.

\section{SIZES IN MASS-LUMINOSITY-FIXED SAMPLES}

\begin{figure}[!t]
\plotone{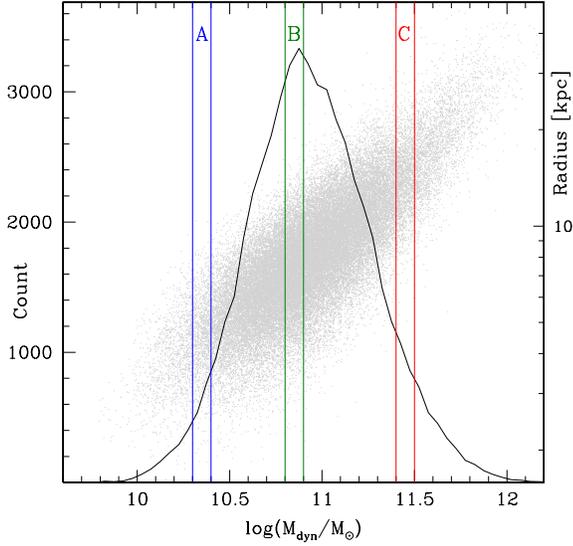}
\caption{ Three sub-samples (A, B and C) defined by using the dynamical mass distribution. The size versus mass relation is displayed in the background (grey dots; right-side ticks) and the galaxy number distribution as a function of the dynamical mass is overlaid (left-side ticks).
\label{samples}}
\end{figure}

\begin{figure}[!t]
\plotone{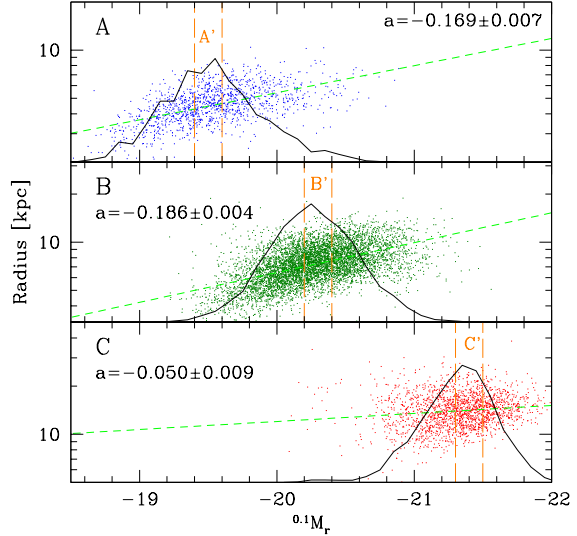}
\caption{ Size versus luminosity relations for the three mass-limited sub-samples. The histograms show the luminosity distribution with normalized Y-axis values. The linear least squares fit is estimated (short-dashed line) and its slope is also denoted in each panel. The vertical long-dashed lines show the luminosity limits for the mass-luminosity-fixed samples (A$'$, B$'$ and C$'$).
\label{lsrel}}
\end{figure}

\begin{figure}[!t]
\plotone{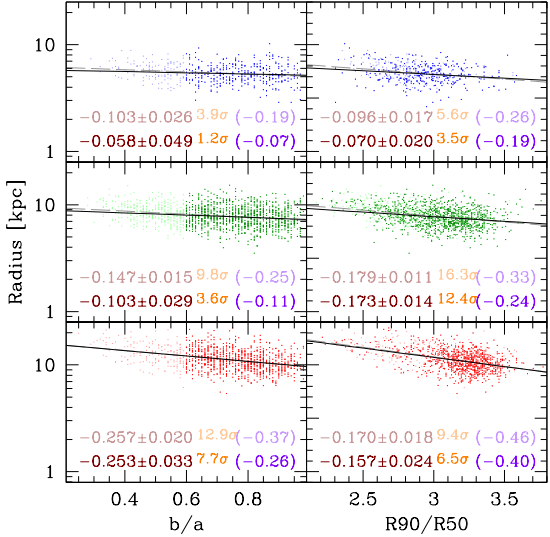}
\caption{ Size versus quantity relations for the three mass-luminosity-fixed samples: A$'$ - upper; B$'$ - middle; and C$'$ - lower panels. The compared quantities are axis ratio (b/a) and light concentration (R90/R50). Light dots show {\it all galaxies} in a given mass-luminosity-fixed sample, while dark dots show only {\it rounder galaxies} with b/a $\ge0.6$.
The linear least squares fits are overlaid (dashed line for {\it all galaxies} and solid line for {\it rounder galaxies}) and their slope$\pm$slope-error's are also denoted in each panel (upper-line for {\it all galaxies} and lower-line for {\it rounder galaxies}). The $\sigma$ value at the upper-right locus of each slope error value indicates the statistical significance of the estimated slope. The numbers in parentheses are the correlation coefficients: the larger absolute value indicates the better correlation.
\label{param0}}
\end{figure}

\begin{figure}[!t]
\plotone{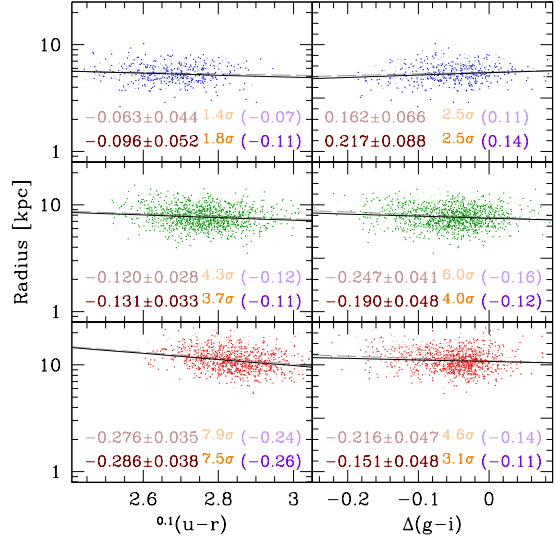}
\caption{ Size versus quantity relations: the same as Figure~\ref{param0}, but the compared quantities are  color ($^{0.1}(u-r)$) and color gradient ($\Delta(g-i)$).
\label{param}}
\end{figure}

Since the influences of mass and luminosity on galaxy size depend on their ranges, we need to test several sub-samples with different masses and luminosities, to understand how the sizes of red early-type galaxies depend on other quantities.
Figure~\ref{samples} shows the selection of three sub-samples based on the dynamical mass distribution: A ($10.3\le\log(M_{\textrm{\scriptsize dyn}}/M_{\odot})\le10.4$; 1299 objects), B ($10.8\le\log(M_{\textrm{\scriptsize dyn}}/M_{\odot})\le10.9$; 6536 objects) and C ($11.4\le\log(M_{\textrm{\scriptsize dyn}}/M_{\odot})\le11.5$; 1926 objects).

Even in the sub-sample within each narrow range of dynamical mass, however, the sizes of red early-type galaxies still depend on their luminosity, as shown in Figure~\ref{lsrel}. The variation of the luminosity-dependency with dynamical mass is reconfirmed in this figure, in the sense that the sizes of the low-mass galaxies (sample A) are more sensitive to luminosity than those of high-mass galaxies (sample C).
To remove the effect of the galaxy size dependence on luminosity, we selected more-confined sub-samples: A$'$ ($-19.6\le$ $^{0.1}M_r\le-19.4$ in A; 301 objects), B$'$ ($-20.4\le$ $^{0.1}M_r\le-20.2$ in B; 1382 objects) and C$'$ ($-21.5\le$ $^{0.1}M_r\le-21.3$ in C; 572 objects). In these sub-samples, the absolute magnitudes and dynamical masses are not significantly correlated with the sizes, because their ranges are sufficiently narrow.
It is noted that our selection criterion for $\sigma_v$ (S/N $\ge20$) rejects many red early-type galaxies in the A$'$ sample ($\sim61\%$), while it rejects only small fractions of galaxies in the B$'$ ($\sim14\%$) and C$'$ ($<1\%$) samples. This implies that the results for the A$'$ sample may be affected by our sample selection unlike those for the B$'$ and C$'$ samples, although it is not clear how such a selection actually affects the results.

Using the three sub-samples fixed in their mass and luminosity (A$'$, B$'$ and C$'$), we first checked the redshift effect in these samples, finding no evidence of the size-redshift relation, which may be due to the narrow redshift range. In addition, we found no evidence of the size-environment relation, using local galaxy number density parameters \citep{lee10}, which agrees with the previous results from literature \citep[e.g.,][]{nai10}.
On the other hand, we found that four quantities show statistically meaningful correlations with the sizes of our sample galaxies: axis ratio (b/a), concentration (R90/R50) \footnote{R90 (R50) is the semimajor axis length of an ellipse containing $90\%$ ($50\%$) of the Petrosian flux.}, color ($^{0.1}(u-r)$) and color gradient ($\Delta(g-i)$), as shown in Figures~\ref{param0} and \ref{param}.
The significance of the correlation is different between the three samples.

In Figure~\ref{param0}, the sizes of galaxies (hereafter, we denote the galaxies without b/a cut as {\it all galaxies}) significantly ($>3\sigma$) depend on their axis ratios in the sense that more elongated galaxies tend to be larger, showing the more significant and strong correlation in the higher-mass sample.
It is noted that about $20-30\%$ of galaxies in each sample have axis ratios smaller than 0.6, which means that they are significantly disky. To investigate the behaviors of the galaxies less affected by their potential disk components, we additionally analyzed the galaxies with b/a $\ge0.6$ only (hereafter, {\it rounder galaxies}). For those {\it rounder galaxies}, the galaxies in higher-mass sample show a stronger size -- axis ratio correlation, similarly to the trend of {\it all galaxies}.

The light concentrations are strongly correlated with the sizes, in the sense that the smaller red early-type galaxies at given mass are more concentrated. This is as expected for the sample galaxies with well-defined radial light profiles (e.g., S{\' e}rsic profiles) in limited ranges of mass and luminosity. This trend is commonly significant for both {\it all galaxies} and {\it rounder galaxies}.

In Figure~\ref{param}, the size dependence on color is not found but the color gradient shows a marginally positive correlation with the size, in the low-mass sample. On the other hand, in the intermediate- and high-mass samples, both the colors and color gradients are clearly correlated with galaxy sizes: larger red early-type galaxies have bluer total colors and bluer outskirts (negative color gradients).
It is noted that the high-mass sample shows stronger trend for the color but weaker trend for the color gradient, compared to the trends in the intermediate-mass sample. Such behaviors do not change even if we limit the sample to {\it rounder galaxies}.

Since the galaxy colors represent their mean stellar ages, the anti-correlation between the size and color index agrees with the result of \citet{sha09} that older galaxies tend to be more compact.
On the other hand, the trends in Figure~\ref{param} disagree with the argument of \citet{nai11} that the scatter in the size-luminosity relation is entirely due to measurement error. Instead, our results show that some physical origins of the scatter exist.

\section{DISCUSSION AND CONCLUSION}

We presented two important results. First, both the masses and luminosities of red early-type galaxies are correlated with their sizes, but differentially at different mass ranges. That is, the sizes of faint galaxies ($\log\, (M_{\textrm{\scriptsize dyn}}/M_{\odot})\simlt10.3$ or $^{0.1}M_{r}\simgt-19.5$) are correlated more significantly with their luminosities, while the sizes of bright galaxies ($\log\, (M_{\textrm{\scriptsize dyn}}/M_{\odot})\simgt11.4$ or $^{0.1}M_{r}\simlt-21.4$) depend more significantly on their dynamical masses.
Second, the correlations of the colors, color gradients and axis ratios with the sizes of red early-type galaxies \emph{at fixed mass and luminosity} depend on their dynamical masses. The sizes of low-mass galaxies ($\log\, (M_{\textrm{\scriptsize dyn}}/M_{\odot})\sim10.35$ and  $^{0.1}M_{r}\sim-19.5$) are relatively less sensitive to the three parameters, while the sizes of intermediate-mass ($\log\, (M_{\textrm{\scriptsize dyn}}/M_{\odot})\sim10.85$ and  $^{0.1}M_{r}\sim-20.3$) and high-mass ($\log\, (M_{\textrm{\scriptsize dyn}}/M_{\odot})\sim11.45$ and  $^{0.1}M_{r}\sim-21.4$) galaxies significantly depend on them.
These results may be related to the different evolutionary processes of red early-type galaxies in different mass ranges.

The sizes of the intermediate- and high-mass red early-type galaxies are correlated with their dynamical masses, colors, color gradients and axis ratios.
\citet{lee08} showed that red early-type galaxies with weak star formation signature tend to have more negative color gradients and smaller axis ratios, compared to purely quiescent red early-type galaxies, which can be interpreted to be the result of recent minor merger events.
If the minor merger affects the colors, color gradients and axis ratios of red early-type galaxies, it may also affect their sizes. Since the minor mergers produce larger fractional changes in size than in mass \citep{ber11a}, a plausible scenario for our results is that the red early-type galaxies with recent minor merger events become slightly larger than the equal-mass counterparts without recent minor mergers.

Another scenario to explain our results for the intermediate- and high-mass red early-type galaxies is that, at fixed mass, rotation causes the large size-to-mass ratio. In this scenario, the small axis ratios (i.e., more elongated shape) of the relatively large galaxies in Figure~\ref{param} may be the result of rotation. We emphasize that this scenario does NOT mean that more massive galaxies tend to rotate faster \citep[it is not true; e.g.][]{cap12}, but does mean that \emph{at fixed mass} rotation makes galaxies larger.
Among early-type galaxies, S0 galaxies are known to be rotating faster than elliptical galaxies \citep{kra11}, and S0 galaxies are thought to be transformed from spiral galaxies via gas stripping \citep[e.g.,][]{kor12}.
Thus, the blue outskirts of the red early-type galaxies with large size-to-mass ratio may be the traces of past young disks in those galaxies.
To confirm which effect is dominant between the minor merger and the rotation, surveys of major-axis long-slit spectroscopy for the intermediate- and high-mass sample galaxies will be useful, which will measure the significance of rotation in those galaxies.

On the other hand, the sizes of low-mass red early-type galaxies are mainly correlated with their luminosities, while the effects of other parameters (mass, color, color gradient and axis ratio) are not significant.
Thus, the minor merger and rotation may not be the main drivers for the size growth of low-mass red early-type galaxies.
Recently, based on the comparison of curvatures in the color-magnitude and color-$\sigma_v$ relations of early-type galaxies, \citet{ber11b} argued that early-type galaxies may have different formation histories depending on their stellar masses: wet mergers dominated at $M^{*} < 3\times10^{10}M_{\odot}$; minor dry mergers dominated at $3\times10^{10}M_{\odot}<M^{*} < 2\times10^{11}M_{\odot}$; and major dry mergers dominated at $M^{*} > 2\times10^{11}M_{\odot}$.
Our results are in agreement with the arguments of \citet{ber11b}, in the sense that the formation of low-mass early-type galaxies does not seem to be dominated by minor dry mergers.
To give robust bases about how the gas-rich mergers make the almost exclusive influence of the luminosities of red early-type galaxies on their sizes, some numerical studies may be additionally required.

Meanwhile, the sizes of the high-mass galaxies show weaker dependence on color gradients but stronger dependence on axis ratios, compared to those of the intermediate-mass galaxies. This seems to be in favor of the scenario in \citet{ber11b} about massive early-type galaxies (major dry merger dominated), because major mergers are expected to make galaxy color gradients shallower \citep{dim09} and may make galaxy shapes more elongated along the direction of collision. However, a major dry merger alone does not cause the excessive growth of a galaxy in size compared to its growth in mass \citep{ber11a}; that is, only major mergers can not explain the size deviation at given mass and luminosity. Thus, for the red early-type galaxies in the high-mass sample (C$'$), the effects of minor and major dry mergers may be combined (e.g., major mergers accompanied by subsequent minor mergers). Otherwise, major merging events may have caused the rotations of massive systems with various angular momenta.

In conclusion, our results provide additional constraints on the different evolution histories of early-type galaxies depending on their masses and luminosities, added on recent results in other studies \citep[e.g.,][]{ber11b}. The sizes of intermediate- and high-mass red early-type galaxies seem to be significantly affected by their recent minor mergers or rotations. On the other hand, the sizes of low-mass red early-type galaxies seem to be the results of some other mechanisms, such as gas-rich major mergers.
Major dry mergers also seem to have influenced on the recent size growth of high-mass red early-type galaxies, but probably combined with subsequent minor mergers or rotation effects. .

\acknowledgments

All authors in Daejeon are the members of Dedicated Researchers for Extragalactic AstronoMy (DREAM) in Korea Astronomy and Space Science Institute (KASI). We are grateful to the anonymous referee for the helpful comments.

\end{document}